# Optical nano artifact metrics using silicon random nanostructures

Tsutomu Matsumoto[1,2], Naoki Yoshida[1], Shumpei Nishio[3], Morihisa Hoga[3], Yasuyuki Ohyagi[3], Naoya Tate[4] & Makoto Naruse[5]

[1] *Graduate School of Environment and Information Sciences, Yokohama National University, 79-7 Tokiwadai, Hodogaya, Yokohama, Kanagawa 240-8501, Japan*

[2] *Institute of Advanced Sciences, Yokohama National University, 79-5 Tokiwadai, Hodogaya, Yokohama, Kanagawa 240-8501, Japan*

[3] *Dai Nippon Printing Co., Ltd., 250-1 Wakashiba, Kashiwa, Chiba 277-0871, Japan*

[4] *Graduate School of Information Science and Electrical Engineering, Kyushu University, 744 Motooka, Nishi-ku, Fukuoka 819-0395, Japan*

[5] *Photonic Network Research Institute, National Institute of Information and Communications Technology, 4-2-1 Nukui-kita, Koganei, Tokyo 184-8795, Japan*

Correspondence and requests for materials should be addressed to T.M.
(email: tsutomu@ynu.ac.jp)

**ABSTRACT**

Nano artifact metrics exploit unique physical attributes of nanostructured matter for authentication and clone resistance, which is vitally important in the age of Internet-of-Things where securing identities is critical. However, high-cost and huge experimental apparatuses, such as scanning electron microscopy, have been required in the former studies. Herein, we demonstrate an optical approach to characterise the nanoscale-precision signatures of silicon random structures towards realising low-cost and high-value information security technology. Unique and versatile silicon nanostructures are generated via resist collapse phenomena, which contains dimensions that are well below the diffraction limit of light. We exploit the nanoscale precision ability of confocal laser microscopy in the height dimension, and our experimental results demonstrate that the vertical precision of measurement is essential in satisfying the performances required for artifact metrics. Furthermore, by using state-of-the-art nanostructuring technology, we experimentally fabricate clones from the genuine devices. We demonstrate that the statistical properties of the genuine and clone devices are successfully exploited, showing that the liveness-detection-type approach, which is widely deployed in biometrics, is valid in artificially-constructed solid-state nanostructures. These findings pave the way for reasonable and yet sufficiently secure novel principles for information security based on silicon random nanostructures and optical technologies.

**Introduction**

Artifact metrics[1], also known as physical unclonable functions (PUFs)[2], use the unique physical properties of ubiquitous materials to improve the information security of devices[3]; this is increasingly important as society enters the age of the Internet of Things[4]. Artifact metrics are based on optical, magnetic, electrical and mechanical properties associated with objects such as ordinary paper[5,6], magnetic microfibers[7], plastics and semiconductor chips[8]. As the number of cloning attacks increases, nano-artifact metrics are growing in importance because they exploit physically uncontrollable processes at the nanoscale. Such techniques are beyond the present roadmap of nanostructuring technologies[9, 10]. In [9], we proposed the concept of *nano*-artifact metrics and experimentally demonstrated the principle by fabricating silicon random nanostructures, which we measured using critical dimension scanning electron microscopy (CD-SEM).

In particular, we use electron-beam (e-beam) lithography to induce the random collapse of photoresists. Resist collapse occurs during the rinsing step of lithography, and it depends on the pattern resolution, resist thickness and duration of e-beam exposure[11]. This leads to the collapse of the intended pattern; the resulting minimum dimensions are smaller than the foreseeable limitations of nanofabrication technologies. In other words, physical randomness on the nanoscale, which has conventionally hindered nanostructuring, provides considerable benefits from the perspective of security. This fundamental principle was demonstrated by evaluating the false match rate (FMR) for verifying identities, false non-match rate (FNMR) for characterising the stability of measurements and the clone-match rate for evaluating the difficulty of making clones[9]. The potential capacity for making unique identities from nanoscale morphologies has been examined theoretically on the basis of eigenanalysis methods[12].

The technology demonstrated so far, however, is difficult to deploy commercially because it requires high-cost equipment with the vacuum environment needed to perform SEM. Although desktop SEMs are currently available[13], the limited operating environments, including safety, hinder their massive deployment.

Optical microscopy suffers from fundamental spatial-resolution limits due to the diffraction of light; thus, performance comparable to an SEM cannot be achieved. However, confocal laser microscopy provides nanoscale resolution in the height dimension, although lateral resolution is limited[14]. Practical optical solutions can be expected if the unique features of a three-dimensional (3D) nanostructure pattern are retrievable via conventional confocal laser microscopy and retain sufficiently secure performance, including authentication and clone resistance.

Herein, we use a conventional confocal microscope to retrieve the patterns of 100 unique, genuine devices comprising silicon random nanostructures. The FMR and FNMR of the genuine devices are investigated as a function of the threshold value for various levels of precision in the height measurements. Nanoscale precision on the order of 1–10 nm in the vertical direction, afforded by confocal microscopy, is essential for obtaining values of the FMR and FNMR that are useful for information security. Five clone devices were fabricated based on original devices and examined using confocal microscopy. Clone resistance was achieved by observing the differences between the height distributions of genuine and clone devices as well as noting a distinct threshold between the error rates of detecting genuine devices and the error rates of detecting clone devices. These results indicate that confocal microscopy is a compatible measurement technique for the secure identification of silicon random nanostructures. The ease of device fabrication and high-throughput rate of characterisation suggest that this is a promising security technology.

**Results and Discussion**

The schematic of the basic architecture of the proposed system is illustrated in Fig. 1a. The random silicon nanostructures were fabricated from an array of pillars, as shown schematically in Fig. 1b. Each pillar had a cross-sectional area of 100 nm × 100 nm and a height of 100 nm; they were positioned on a 200 nm × 200 nm square grid that filled a 20 μm × 20 μm square area. To facilitate alignment, a 30 μm × 30 μm square frame and an alignment mark were drawn outside the pillar-array area. We used an e-beam lithography system (JEOL JBX-9300FS) with a 100-kV acceleration voltage. The rinsing step induced the random collapse of resist pillars. The resulting pattern was captured by CD-SEM (Hitachi High-Technologies CG4000) and the random structures are shown in Fig. 1c. We constructed 100 unique devices on a single wafer using these pillar-array structures.

The fabricated patterns were observed with a confocal laser microscope (Olympus Inc. LEXT OLS400). This step corresponds to the '**Scan$_A$**' stage in Fig. 1a; an example image is shown in Fig. 1d. An area of 126 × 126 pixels at the centre of the device was extracted; each pixel corresponded to a square area that was 125 nm on a side (~15 μm$^2$). The original data was recorded during the process '**Scan$_R$**' in Fig. 1a, which became the template for the decision process used during authentication.

During the '**Decision**' process in Fig. 1a, we used the statistical properties of the optical height measurements of a genuine device to detect and reject clone devices. As shown in Fig. 1d, the pattern of a genuine device contained a large amount of height information. We cloned the patterns of original devices using state-of-the-art, silicon-nanostructure technologies. As a result, the statistical properties of the optical measurements of clones differed significantly from those of genuine devices, as we demonstrate later below. This approach is similar to the liveness detection utilised in biometrics[15].

If an optical measurement, which we denote by $A(i,j)$, passes through the initial screening process stated above, we calculate the correlations between the device under study and the template stored in the system, denoted $B(i,j)$, based on the Pearson correlation coefficient:

$$R = \frac{\sum_{i,j}\left[A(i,j)-\overline{A}\right]\left[B(i,j)-\overline{B}\right]}{\sqrt{\sum_{i,j}\left[A(i,j)-\overline{A}\right]^2}\sqrt{\sum_{i,j}\left[B(i,j)-\overline{B}\right]^2}}, \qquad (1)$$

where $\overline{A}$ and $\overline{B}$ are the averages of $A(i,j)$ and $B(i,j)$, respectively. Moreover, the image is shifted between one and three pixels to the upper, lower, left and right sides; $R$ is calculated for each shifted position. The maximum value of $R$ from these positions is used to quantify the similarity between $A(i,j)$ and $B(i,j)$.

Based on $R$, the FMR and FNMR are evaluated as indicators of individuality and measurement stability, respectively. All 100 devices are used to calculate the FMR. If the similarity value given by Eq. (1) is greater than a given threshold, the two images are presumed to be the same, which is false. To evaluate the FNMR, each of the 100 samples is measured 10 times. If the similarity between two observations of the same sample is smaller than a given threshold, the two images are considered to be different devices, which is false. The solid blue curve in Fig. 2 shows the FMR, which remains small even at low values of the threshold around 0.1. The solid red curve depicts the FNMR, which increases with the threshold for values larger than 0.5. These results indicate that it is possible to obtain sufficiently small values of the FMR and FNMR by choosing an appropriate threshold.

We further evaluate the FMR and FNMR by intentionally degrading the resolution of the height measurements. Specifically, the measurement data is rounded off at the 1-nm, 10-nm and 100-nm levels of precision and the resulting FMR are represented by the dotted, dashed and dot-dashed curves, respectively, in Fig. 2 (the same depictions are also used for the FNMR). The performance of 1-nm- and 10-nm-resolution measurements nearly overlaps with the original data, whereas the 100-nm-resolution measurement deviates from the original; in particular, the FNMR is significantly degraded. This result confirms that the security performance of this technique derives from the nanoscale height characteristics of the silicon random nanostructures.

Attackers can perform optical measurements on the genuine devices, so we fabricate five different kinds of clone devices based on their parent devices and examine the possibilities of detecting and rejecting cloning attacks. The experimental process of clone fabrication is as follows: Based on measurements obtained from the confocal microscope, we design the structure for the clone device. Binary-level height data may be used for most of the clone devices; in some instances, multiple height levels may be used. This degree of accuracy is reasonable given the nanostructure technologies that will be accessible to attackers in the foreseeable future concerning especially the fact that fabricating multiple-level nanostructures containing complex morphology is extremely difficult in so-called grey-scale e-beam lithography[16,17] and multiexposure methods[18]. In this study, we binarise the measurement data and fabricate binary-level patterns using e-beam lithography; then we use atomic force microscopy (AFM) to confirm that the fabrication is successful. We used an Olympus OLS3500 atomic force microscope to examine the surface profiles of the fabricated clones using a high-aspect ratio tip cantilever with a half-tip angle of 6° (Olympus OLCL-AC160BN-A2) and a lateral resolution of 39 nm. An AFM image of a fabricated clone device and its laser microscopy image are shown in Figs. 3a and 3b, respectively. About a quarter of the whole area of the device is depicted in Figs. 3a and 3b for the sake of presenting a magnified top-down view of the structures. Figure 3a confirms that the binary-level surface profile of the clone is executed precisely. Figure 3b shows that the optical measurement retrieves the surface profile of the clone. It is noteworthy that the optical height measurements do not exhibit binary levels, which may be attributed to diffraction and scattering induced by laser microscopy. We find that the statistical distributions of height vary considerably for genuine and clone devices. Genuine devices exhibit a Gaussian-like distribution (Fig. 3c), whereas clones exhibit skewed statistics (Fig. 3d); this property can be exploited for detecting clones.

We evaluate similarities between the height distributions using Eq. (1) and replacing $A(i, j)$ and $B(i, j)$ with $H_A(i)$ and $H_B(i)$, respectively, where $H_A(i)$ and $H_B(i)$ are the height histograms of $A(i, j)$ and $B(i, j)$, respectively. If both $H_A(i)$ and $H_B(i)$ originate from genuine devices, the similarity must be larger than a certain threshold to indicate that both belong to the same category (genuine device); this means that the inspection is successful (no error). On the other hand, if either $H_A(i)$ or $H_B(i)$ originates from a clone device, the calculated value of the similarity is smaller, indicating the detection of a clone. The solid blue curve in Fig. 3e shows the error ratio for the similarities between the 100 different genuine devices; the dashed red curve depicts the error ratio between the five clone devices and the 100 genuine devices. Notably, the former and the latter curves cluster on the left (less than 0.2) and right (greater than 0.27), respectively; thus, the detection of clones is possible by setting the threshold to be an intermediate value.

In conclusion, we experimentally demonstrate an optical approach for using the unique nanoscale fingerprints of silicon random structures to realise low-cost nano-artifact metrics produced in a vacuum-free environment. Specifically, we experimentally validate that conventional confocal laser microscopy can perform precise, nanoscale measurements that accommodate the individuality and measurement stability of silicon random nanostructures. We experimentally fabricate clone devices based on parent devices and show that clone resistance can be achieved by inspecting the statistical attributes of the genuine and clone devices; this demonstrates that the liveness-detection-type approach, which has been successfully employed in biometrics, is validated in artificially-constructed solid-state nanostructured devices.

**Acknowledgements**

This work was supported in part by the Grant-in-aid in Scientific Research and the Core-to-Core Program, A. Advanced Research Networks from the Japan Society for the Promotion of Science.

**Author Contributions**

T.M. directed the project; T.M., M.N., N.T. and M.N. designed system and device architectures; S.N., M.H. and Y.O. performed experimental device fabrication; N.Y., S.N. and M.H. performed data acquisition; N.Y. and M.N. conducted data analysis and visualisation; and T.M., N.Y and M.N. wrote the paper.

**Additional Information**

**Competing financial interests:** The authors declare no competing financial interests.

# References

1. Matsumoto, H. & Matsumoto, T. Clone match rate evaluation for an artifact-metric system. *IPSJ J.* **44,** 1991-2001 (2003).

2. Pappu, R., Recht, B., Taylor, J. & Gershenfeld, N. Physical one-way functions. *Science* **297,** 2026-2030 (2002).

3. van Renesse, R. L. *Optical document security* (Artech House, Boston, 2004).

4. Gubbi, J., Buyya, R., Marusic, S. & Palaniswami, M. Internet of Things (IoT): A vision, architectural elements, and future directions. *Future Generation Computer Systems* **29**, 1645-1660 (2013).

5. Buchanan, J. D. R. *et al*. Forgery: 'Fingerprinting' documents and packaging. *Nature* **436,** 475 (2005).

6. Yamakoshi, M., Tanaka, J., Furuie, M., Hirabayashi, M. & Matsumoto, T. Individuality evaluation for paper based artifact-metrics using transmitted light image. *Proc. SPIE* **6819,** 68190H (2008).

7. Matsumoto, H., Takeuchi, I., Hoshino, H., Sigahara, T. & Matsumoto, T. An artifact-metric system which utilizes inherent texture. *IPSJ J*. **42,** 139-152 (2001).

8. Lim, D. *et al.* Extracting secret keys from integrated circuits. *IEEE Trans. VLSI Syst.* **13,** 1200 (2005).

9. Matsumoto, T. *et al.* Nano-artifact metrics based on random collapse of resist. *Sci. Rep.* **4,** 6142 (2014).

10. Naruse, M., Tate, N. & Ohtsu, M. Optical security based on near-field processes at the nanoscale. *J. Opt.* **14,** 094002 (2012).

11. Namatsu, H., Kurihara, K., Nagase, M., Iwadate, K. & Murase, K. Dimensional limitations of silicon nanolines resulting from pattern distortion due to surface tension of rinse water. *Appl. Phys. Lett.* **66,** 2655-2657 (1995).

12. Naruse, M. *et al.* Eigenanalysis of morphological diversity in silicon random nanostructures formed via resist collapse. *arXiv* 1602.08205 (2016). http://arxiv.org/abs/1602.08205

13. Lawrance, J., Carruthers, J., Jiao, J., & Berger, S. Comparison of Materials Characterization Performed by Low Voltage Desktop SEM and Standard High Resolution SEM. *Microscopy and Microanalysis* **13,** 1728-1729 (2007).

14. Minsky, M. Memoir on inventing the confocal scanning microscope. *Scanning* **10,** 128-138 (1988).

15. Nixon, K. A. *et al.* Novel spectroscopy-based technology for biometric and liveness verification. *Proc. SPIE* **5404,** 287-295 (2004).

16. Schleunitz, A. & Schift, H. Fabrication of 3D nanoimprint stamps with continuous reliefs using dose-modulated electron beam lithography and thermal reflow. *J. Micromech. Microeng.* **20,** 095002 (2010).

17. Kurihara, M. Status and Challenges of Template Fabrication Process for UV Nanoimprint Lithography. *J. Photopolymer Sci. Tech.* **20,** 563-567 (2007).

18. del Campo, A. & Greiner, C. SU-8: a photoresist for high-aspect-ratio and 3D submicron lithography. *J. Micromech. Microeng.* **17,** R81 (2007).

**Figure legends**

**Figure 1. Architecture for optical nano-artifact metrics based on silicon random nanostructures.** (a) Fundamental system architecture of nano-artifact metrics comprising optical measurements and decision processes. Conventional confocal laser microscopy is employed to exploit its nanoscale height resolution as well as to utilise the intrinsic attributes of silicon random nanostructures. (b) Schematic of an original array of pillars prior to its intentional collapse during the rinsing step of e-beam lithography; the collapsed array is a versatile, three-dimensional (3D) nanostructure. (c) SEM image of a fabricated silicon nanostructure. (d) Confocal laser microscope image of the device obtained with lateral resolution of 125 nm and nanoscale height resolution.

**Figure 2. Evaluation of security performance.** False match rate (FMR) and false non-match rate (FNMR) as a function of threshold; they are evaluated to verify device individuality and measurement stability, respectively. The FMR and FNMR are also calculated while intentionally degrading the height resolution by rounding the original data in 1-nm, 10-nm and 100-nm levels of precision; the result confirms that nanoscale information is responsible for the security performance of the devices.

**Figure 3. Clone rejection.** Based on the optical measurements of genuine devices, we experimentally fabricate *clone* devices using e-beam lithography. (a) AFM image and (b) Confocal microscopy image of a fabricated clone device. The statistical characteristics of optically acquired height information for (c) genuine and (d) clone devices. (e) Clone

rejection can be achieved by evaluating the statistical similarity between the height distributions of the genuine and clone devices.

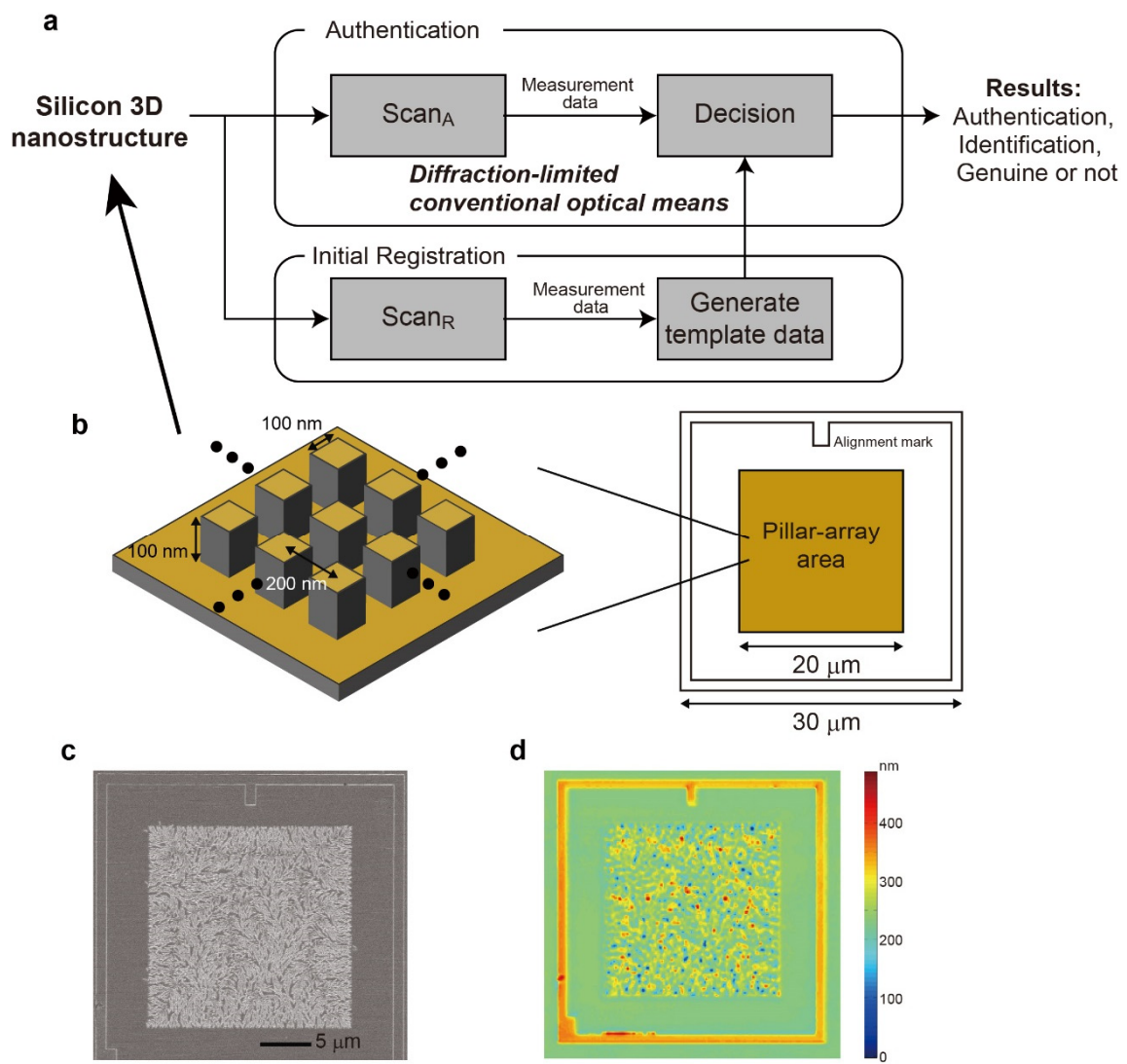

Figure 1

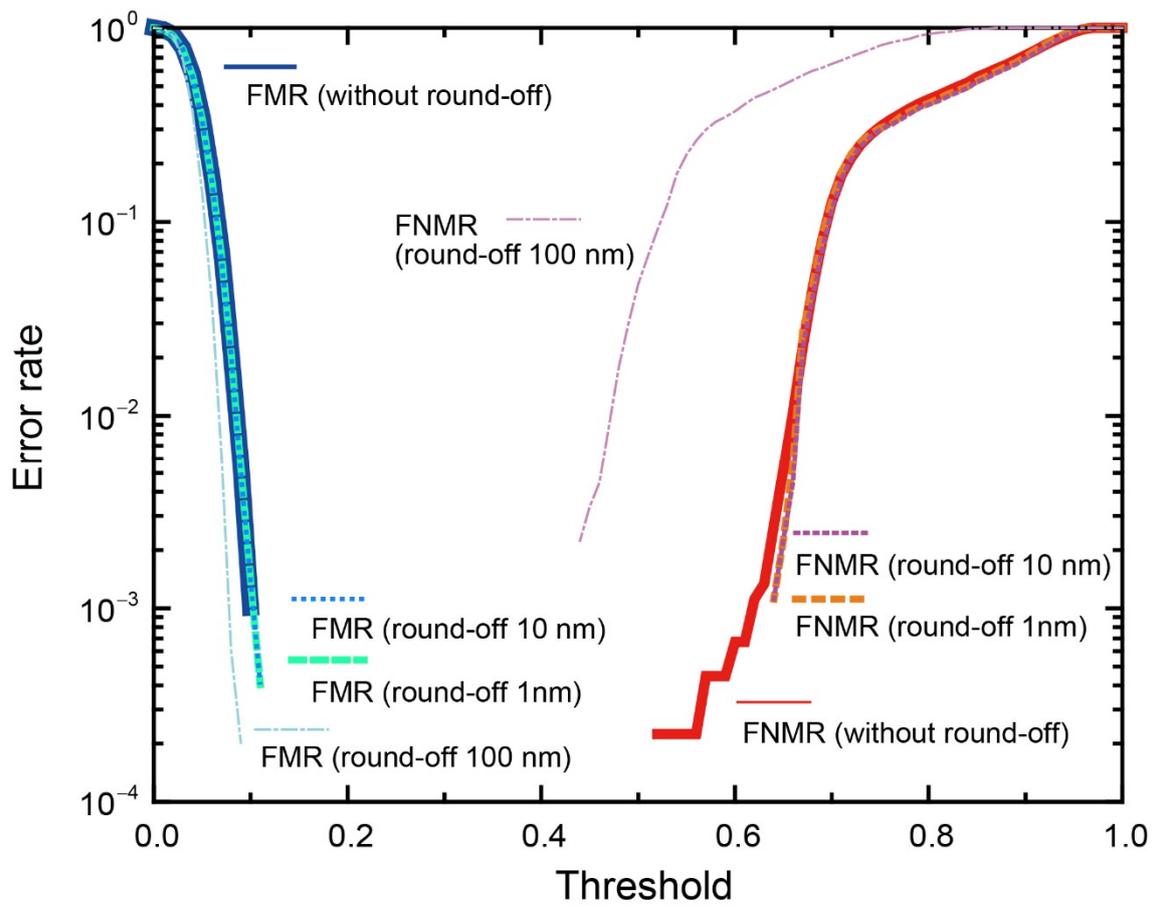

Figure 2

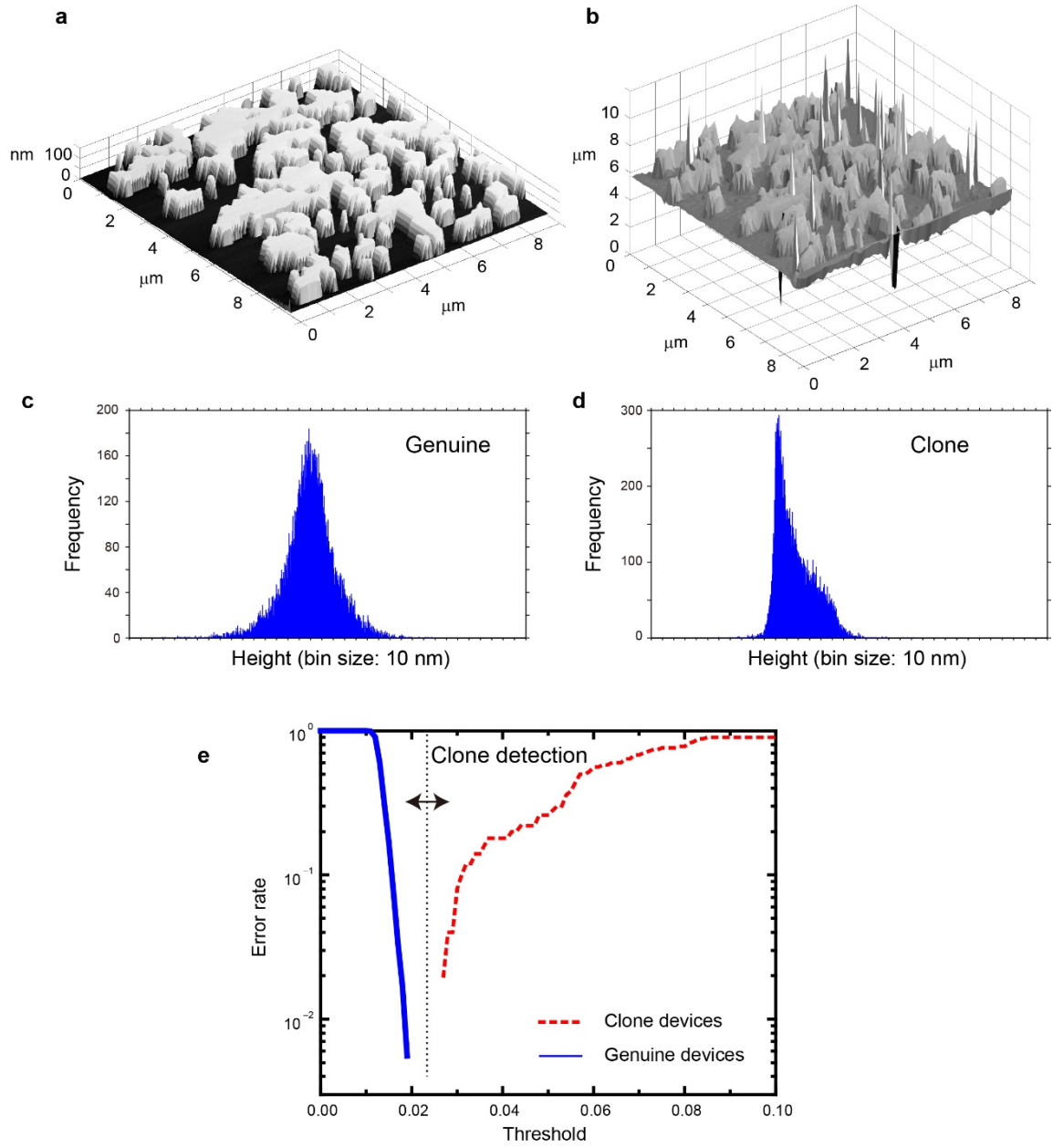

Figure 3